\documentstyle[amssymb,twocolumn,aps,prl]{revtex}

\begin{document}
\title{{\bf Intrinsic accuracy in 3-dimensional photoemission band mapping }}
\author{V. N. Strocov\cite{IHPC}}
\address{Experimentalphysik II, Universit\"{a}t Augsburg, D-86135 Augsburg,\\
Germany}
\date{\today }
\maketitle

\begin{abstract}
Fundamental principles of mapping 3-dimensional quasiparticle dispersions in
the valence band using angle-resolved photoemission spectroscopy are
discussed. Such mapping is intrinsically limited in accuracy owing to
damping of the final states, resulting in equivalent broadening in the
surface-perpendicular wavevector. Mechanisms of the intrinsic accuracy are
discussed in depth based on a physically transparent picture involving
interplay of the final- and initial-state spectral functions, and
illustrated by photoemission simulations and experimental examples. Other
interesting effects of 3-dimensional dispersions include 'ghost'
photoemission peaks outside the Fermi surface and finite peak width at the
Fermi level. Finally, optimization of the experiment on the intrinsic
accuracy is discussed.
\end{abstract}

\bigskip


\section{Introduction: Why intrinsic accuracy?}

The quasiparticle bandstructure $E({\bf k})$, which reflects the peaks of
the spectral function $A(E,{\bf k})$ depending on energy and wavevector $%
{\bf k}$, is the key property of the crystalline solids. Angle-resolved
photoemission (PE) spectroscopy (for reviews see, e.g., \cite%
{Kevan92,Hufner95}) is the main experimental method to map $E({\bf k})$ with
resolution in energy $E$ and, in principle, in the 3-dimensional ${\bf k}$ (%
{\em 3D band mapping}). This is based on conservation of ${\bf k}$ in the
process of photoexcitation from the occupied initial state to unoccupied
final state in the bulk of the crystal. However, the PE experiment is
inherently performed on the crystal terminated by the surface. By virtue of
the remaining surface-parallel periodicity, the surface-parallel wavevector $%
{\bf k}_{\parallel }$ is conserved in the whole system and can be determined
from the surface-parallel vacuum wavevector ${\bf K}_{\parallel }$ with the
accuracy limited only by instrumental resolution. As the
surface-perpendicular periodicity is however broken, the
surface-perpendicular wavevector $k_{\perp }$ is not conserved in the whole
system and can not be directly determined. Control over $k_{\perp }$ is
therefore a fundamental problem of 3D band mapping.

The control over $k_{\perp }$ has two aspects. First, $k_{\perp }$ is
distorted when the photoelectrons exit from the bulk of the solid into
vacuum. As $k_{\perp }$ is conserved in the very photoexcitation process in
the bulk, it can nevertheless be recovered if the $k_{\bot }$-dispersion of
the final state $E^{f}(k_{\bot })$ or that of the initial state $%
E^{i}(k_{\bot })$ back in the bulk is known. Commonly an empirical
free-electron-like approximation is used here for the final states, though
the accuracy of this can often be insufficient. Nevertheless, the true
quasiparticle $E^{f}(k_{\bot })$ including the non-free-electron and
self-energy effects in the final state can be determined using
very-low-energy electron diffraction (VLEED) spectroscopy (see \cite%
{Exi,TMDCs,CFS} and references therein) based on that the PE final states,
as proved by the one-step theory of PE, are the time-reversed LEED states %
\cite{Feibelman74,Pendry76}. It is a common point of view that knowledge of
the final-state $E({\bf k})$ exhaust the whole problem of 3D band mapping.
Indeed, variation of the final-state energy $E^{f}$ in the PE experiment
(commonly achieved by variation of the photon energy $h\nu $) would through
known $E^{f}(k_{\bot })$ be translated into variations of $k_{\bot }$, and
accurate 3D mapping of the occupied $E({\bf k})$ would be performed.

There is another aspect however. Due to inelastic absorption and elastic
reflection from the crystal potential the PE final state is intrinsically
damped towards interior of the solid \cite{Feibelman74,Pendry76}. Such a
confinement in the surface-perpendicular coordinate is equivalent, by the
uncertainty principle, to certain {\em intrinsic broadening in }$k_{\perp }$%
. The PE signal via this final state is then formed by all initial states
within the $k_{\perp }$-broadening interval, and the PE peak reflects an
average of the quasiparticle $E^{i}(k_{\bot })$ in the valence band. A
difference of the measured averaged $k_{\perp }$-dispersion from the true $%
E^{i}(k_{\bot })$ limits {\em intrinsic accuracy} of 3D band mapping. It is
borne by fundamental physics of the PE process, and can not be improved
instrumentally.

The intrinsic accuracy appears thus as an {\em extrinsic} factor to the
excited-state self-energy effects as deviations of the quasiparticle
excited-state $E({\bf k})$, measured in the ideal PE experiment with
undamped final states, from the ground-state bandstructure.

The intrinsic accuracy problem has in pieces been tackled in quite a few
papers. Here, my aim is to delineate an entire concept, including essential
physics and mechanisms of the intrinsic accuracy, and unveil its non-trivial
effects encountered in the PE experiment. The concept is built upon a simple
picture involving interplay of the final- and initial-state spectral
functions, which gives a clearer physical insight compared to the explicit
one-step PE theory with its heavy computational machinery. Finally, I
discuss optimization of the PE experiment with respect to the intrinsic
accuracy of 3D band mapping.

\section{Basic physics of the PE process}

In the following analysis of the PE process it is implied that $E({\bf k})$
is the quasiparticle bandstructure, already incorporating the excited-state
self-energy effects. We will neglect the surface effects and concentrate on
the bulk derived states with significant dispersion in $k_{\perp }$
(so-called {\em 3D states}). The ideal $E$- and ${\bf k}_{\parallel }$%
-resolution of the PE experiment is assumed.

\subsection{Initial and final states of PE}

First we will recall physics of the PE initial and final states based on the
ideas from \cite{Kevan92,Hufner95,Matzdorf96}. For simplicity we will assume
that their wavefunctions have only one Bloch wave constituent.

The final-state wavefunction is the time-reversed LEED wavefunction. As
sketched in Fig.1 (the upper panels represent the final state), the
wavefunction is described by a Bloch wave which is damped in the
surface-perpendicular direction $r_{\bot }$ towards the crystal interior %
\cite{Exi,CFS,Pendry69,Pendry74}. There are two sources of such damping: 1)
inelastic absorption, expressed by absorption potential $V_{i}$ connected
with the electron lifetime as $2V_{i}=$ $\hbar /\tau _{e}$, and 2) elastic
reflection from the crystal potential in the final-state band gaps. The
damping is described by complex $k_{\bot }^{f}$ component of the final-state
wavevector ${\bf k}^{f}$, with $Imk_{\bot }^{f}$ connected to the
photoelectron escape length $\lambda $ as $2Imk_{\bot }^{f}=\lambda ^{-1}$
(the factor of 2 comes from squaring the wavefunction amplitude for
intensity).

The corresponding final-state dispersion $E^{f}(k_{\bot })$, shown in Fig.1
by black lines, is radically different from this one if the Bloch waves were
undamped (propagating), as shown in gray: the dispersions as a function of $%
Rek_{\bot }^{f}$ are smooth and {\em gapless}, passing continuously through
the band gaps of the undamped $E({\bf k})$ \cite{Exi,CFS,Pendry69,Pendry74}.
Such regions will in the following be somewhat loosely referred to as
final-state band gaps. Note that due to surface-parallel invariance of the
PE process the Bloch waves are undamped in the surface-parallel direction,
retaining real ${\bf k}_{\parallel }$ and gapped ${\bf k}_{\parallel }$%
-dispersions \cite{CFS}.

The final-state spectral function $A^{f}(E,{\bf k})$, due to the finite
photoelectron lifetime, is in principle characterized by certain
distribution in $E$. However, in the PE experiment this is reduced to the $%
\delta $-function at the energy given by the PE\ analyser. Due to damping of
the final-state wavefunction $A^{f}(E,{\bf k})$ is characterized, on the
contrary, by Lorentzian distribution in real $k_{\bot }$, as shown in Fig.1,
centered at $k_{\bot }^{0}=Rek_{\bot }^{f}$ of the final-state Bloch wave
and having the fullwidth $\delta k_{\bot }=2Imk_{\bot }^{f}$ (in the
angle-resolved PE experiment ${\bf k}_{\parallel }$ is fixed and can be
omitted). Therefore, $A^{f}(E,{\bf k})$ is given by 
\begin{equation}
A^{f}(k_{\bot })\propto \frac{\delta k_{\bot }}{\left( k_{\bot }-k_{\bot
}^{0}\right) ^{2}+(\delta k_{\bot }/2)^{2}}
\end{equation}%
where the nominator stands for the normalization $\int\limits_{-\infty
}^{+\infty }A^{f}(k_{\bot })dk_{\bot }=1$. The final state of PE\ is thus
characterized by {\em fixed }$E${\em \ and broadening in }$k_{\bot }$ whose
width is determined by $Imk_{\bot }^{f}$ of the damped final-state Bloch
wave \cite{Kevan92,Hufner95,Matzdorf96}.

Nature of the PE initial state is different and complementary. The
initial-state wavefunction is basically undamped: Removal of holes by the
finite hole lifetime is compensated by their simultaneous generation by the
electromagnetic field, which is almost homogeneous, extending into the
crystal interior over a light absorption length of $\sim $100 \AA , very
large compared to the final-state damping length. As sketched in Fig.1 (the
lower panels represent the initial state), initial-state wavefunction is
therefore described by an almost undamped Bloch wave, having well-defined
real $k_{\bot }$.

The corresponding initial-state dispersion $E^{i}(k_{\bot })$ shows up sharp
and {\em gapped} $k_{\bot }$-dispersions typical of the undamped states,
radically different from the final-state.

The initial-state spectral function $A^{i}(E,{\bf k})$ is characterized,
owing to well-defined $k_{\bot }$, by distribution in real $k_{\bot }$
reduced to the $\delta $-function. Due to finite hole lifetime $\tau _{h}$
it is characterized, on the contrary, by Lorentzian distribution in energy,
as shown in Fig.1, centered at the band energy $E^{i}(k_{\bot })$ and having
the fullwidth $\delta E=\hbar /\tau _{h}$. Therefore, $A^{i}(E,{\bf k})$ is
given by%
\begin{equation}
A^{i}(E)\propto \frac{\delta E}{\left( E-E^{i}(k_{\bot })\right) ^{2}+\left(
\delta E/2\right) ^{2}}
\end{equation}%
where the nominator ensures $\int\limits_{-\infty }^{+\infty }A^{i}(E)dE=1$.
The initial state of PE is thus characterized, complementarily to the final
state, by {\em fixed }$k_{\perp }${\em \ and broadening in }$E$ whose width
is determined by the hole lifetime \cite{Kevan92,Hufner95,Matzdorf96}.

Note that some implementations of the one-step PE theory \cite%
{Pendry76,Nilsson80} describe the hole lifetime effects by damping of the
initial states. However, such an approach is in clear contradiction with the
PE experiment, which unambiguously finds in the valence band the gapped $%
E^{i}(k_{\bot })$ dispersions typical of undamped states. The same
fundamental flaw is implied by an PE lineshape analysis (see, e.g., \cite%
{Knapp79,Smith93}) where the $E$-broadening of the initial state is replaced
by an equivalent $k_{\perp }$-broadening $\sim \delta E\left( \frac{\partial
E^{i}}{\partial k_{\bot }}\right) ^{-1}$. Although such a replacement does
not alter the lineshape while the $k_{\bot }$-dispersions remain linear, it
leaves out some of the intrinsic effects discussed here.

Inverse photoemission (IPE) is commonly treated as the time-reversed PE
process \cite{Pendry80} in which the IPE final state in the conduction band
is equivalent to the PE initial state in the valence band. However, the
above arguments suggest a fundamental difference between the two: as the
IPE\ final state is excited by the damped LEED wavefunction, it is, in
contrast to the PE\ initial state, also damped. Correspondingly, the
conduction band $E({\bf k})$ measured in the IPE experiment should be
characterized by gapless $k_{\perp }$-dispersions similar to the PE
final-state dispersions in Fig.1. Surprisingly, such a fundamental
difference between the IPE and PE processes, tracing back to fundamental
principles of quantum mechanics, has so far been addressed nor theoretically
neither experimentally.

\subsection{Development of the PE spectrum: Intrinsic $k_{\bot }$-resolution}

We will now apply the above principles to analyze the PE response of 3D
states. We will use a simple formalism based on interplay of the spectral
functions \cite{Kevan92,CFS,Matzdorf96}, which is extremely simple and gives
a clear physical insight. Its accuracy, despite principal limitations
compared to the explicit one-step PE theory (see \ref{OneStep}), is
nevertheless sufficient for evaluation of the PE peak profiles \cite{Berge02}%
.

As illustrated in Fig.2, the photoemission current $I(E^{f},E^{i})$ at the
final-state energy $E^{f}$ and initial-state energy $E^{i}=E^{f}-h\nu $ is
formed by adding up the elementary photocurrents $dI(E^{f},E^{i})$ from all $%
dk_{\bot }$ intervals, in which direct transitions take place, with $k_{\bot
}$ spanning the whole Brillouin zone (BZ). Each $dI(E^{f},E^{i})$ is
weighted by the above spectral functions, $A^{f}(k_{\bot })$ to express
closeness of $k_{\bot }$ to $k_{\bot }^{0}$ of the final-state Bloch wave,\
and $A^{i}(E^{i})$ to express closeness of $E^{i}$ to the initial-state band
energy $E^{i}(k_{\bot })$ at this $k_{\bot }$ with $\delta E$ varying along $%
E^{i}(k_{\bot })$. Moreover, each $dI(E^{f},E^{i})$ is multiplied by
amplitude factors of the final-state surface transmission $|T^{f}|^{2}$ and
photoexcitation matrix element $\left| M_{fi}\right| ^{2}$. This yields $%
dI(E^{f},E^{i})=dk_{\bot }\left| T^{f}\right| ^{2}\left| M_{fi}(k_{\bot
})\right| ^{2}A^{f}\left( k_{\bot }\right) A^{i}\left( E^{i}\right) $. To
obtain the whole photocurrent, this expression should be integrated in $%
k_{\bot }$: 
\begin{eqnarray}
I(E^{f},E^{i}) &\propto &\int\limits_{-\infty }^{+\infty }dk_{\bot }\left|
T^{f}\right| ^{2}\left| M_{fi}(k_{\bot })\right| ^{2}\cdot  \label{IntF} \\
&&\frac{\delta k_{\bot }}{\left( k_{\bot }-k_{\bot }^{0}\right) ^{2}+(\delta
k_{\bot }/2)^{2}}\cdot  \nonumber \\
&&\frac{\delta E}{\left( E^{i}-E^{i}(k_{\bot })\right) ^{2}+(\delta E/2)^{2}}
\nonumber
\end{eqnarray}%
This integral can be evaluated at negligible computational cost (it can even
be performed analytically using Taylor expansion of $E^{i}(k_{\bot })$ \cite%
{CFS}).

Cuts of $I(E^{f},E^{i})$ along the $E^{f}=const$ line correspond to the
constant-final-state (CFS) measurement mode, along the $h\nu
=E^{f}-E^{i}=const$ line to the energy-distribution-curve (EDC) mode, and
along $E^{i}=const$ to the constant-initial-state (CIS) mode. Fig.2
illustrates, for simplicity, the PE spectrum taken using the CFS mode in
which $k_{\bot }^{0}$ is constant throughout the spectrum.

The PE spectrum shows up, whichever the measurement mode, a peak centered
roughly at the energy $E^{i}(k_{\bot }^{0})$ dictated by the direct
transition at $k_{\bot }^{0}$ between the final and initial bands.
Broadening of the peak results from the final-state $k_{\bot }$-broadening
combined with the initial-state $E$-broadening. In this context the $k_{\bot
}$-broadening appears as an {\it intrinsic }$k_{\bot }${\it -resolution} of
the PE experiment (in the following we will use these terms interchangeably
depending on the context). Note that this resolution, in contrast to the
instrumentally limited resolution in ${\bf k}_{\parallel }$, is limited by
intrinsic mechanisms involved in the PE process and can not be improved
instrumentally.

\subsection{Regimes of the PE experiment}

\label{Regimes}If the $k_{\perp }$-resolution is at its best extreme $\delta
k_{\bot }\longrightarrow 0$, then $A^{f}\left( k_{\bot }\right) $ becomes $%
\delta $-function $\delta (k_{\bot }-k_{\bot }^{f})$ and the integral (\ref%
{IntF}) for $I(E^{f},E^{i})$ is reduced to $A^{i}\left( E\right) $ whose
maximum points to the initial band energy $E^{i}(k_{\bot }^{0})$ in the $%
{\bf k}$-point dictated by the direct transition. The PE peaks will then
exactly follow the 3D quasiparticle dispersions in the valence band. In
practice, such a situation is realized under the condition%
\begin{equation}
\delta k_{\bot }\ll k_{\bot }^{BZ}  \label{BSRegime}
\end{equation}%
where $k_{\bot }^{BZ}$ is the surface-perpendicular dimension of the BZ.
This condition identifies the so-called {\em bandstructure regime}
(BS-regime) of the PE experiment which ensures accurate 3D band mapping \cite%
{Hufner95,Feibelman74}. In the real space, it reads as $\lambda \gg c_{\perp
}$, where $c_{\perp }$ is the surface-perpendicular dimension of the unit
cell.

If the PE final state comprise a few Bloch waves, the condition (\ref%
{BSRegime}) should be generalized as $\delta k_{\bot }\ll \frac{k_{\bot
}^{BZ}}{N}$, where $N$ is the number of the corresponding final bands. Often
in the literature (see, e.g., \cite{Feibelman74}) this number is confused
with the number of {\em all} unoccupied bands available for given $E^{f}$
and ${\bf k}_{\parallel }$; as the latter becomes with energy progressively
immense, an erroneous conjecture is made that at high energies any 3D band
mapping is impossible. However, in the multitude of the available unoccupied
bands only those are effective as the final bands, whose Bloch waves provide
effective transport of the photoelectrons out of the solid by significant
coupling to the outgoing plane wave \cite{CFS,CFC-Vg}. Typically there are
only a few ($N=1-2$) final bands in this sense (see examples in \ref%
{Non-linearity}).

If the $k_{\perp }$-resolution is at the opposite extreme $\delta k_{\bot
}\longrightarrow \infty $, any resolution in $k_{\bot }$ is lost, $%
A^{f}\left( k_{\bot }\right) $ becomes constant and the integral (\ref{IntF}%
) yields, assuming that $|T^{f}|^{2}$ and $\left| M_{fi}\right| ^{2}$ are
independent of $k_{\bot }$ and $E^{i}(k_{\bot })$ is linear, the
1-dimensional DOS (1DOS) $\frac{\partial k_{\bot }}{\partial E^{i}}$. The PE
peaks will then manifest the 1DOS maxima independently of $k_{\bot }^{0}$.
In practice, such a situation is achieved under the condition $\delta
k_{\bot }\gtrsim k_{\bot }^{BZ}$, which identifies the so-called {\em %
1DOS-regime} \cite{Hufner95,Feibelman74}.

Interpretation of the PE data between the bandstructure and 1DOS regimes is
doubtful \cite{Feibelman74,Grobman75}: the PE peaks retain some $h\nu $%
-dispersion, although damped by strong averaging in $k_{\bot }$, but they
can be far from the true quasiparticle $E^{i}(k_{\bot })$.

Evolution of the regimes of the PE experiment with the $k_{\perp }$%
-resolution will now be illustrated by a PE simulation. The model employs
the initial state whose parameters are chosen to resemble roughly the {\it sp%
}-band of Cu \cite{CFS}: It is described by the free-electron dispersion $%
E^{i}(k_{\bot })=\frac{\hbar ^{2}}{2m}k_{\bot }^{2}+V_{000}$ and the energy
broadening $\delta E$ varying linearly as a function of energy to match zero
at the Fermi level $E_{F}$ and an experimental value of 1.7 eV at the bottom
of the band. The Fermi-Dirac cut is introduced as the ideal step function.
The photocurrent $I(E^{f},E^{i})$ is evaluated by integration of the
spectral functions in $k_{\bot }$ (\ref{IntF}), assuming constant amplitude
factors $|T^{f}|^{2}$ and $\left| M_{fi}\right| ^{2}$. The integration also
extends above the Fermi vector $k_{\bot }^{F}$ to include certain
contribution from the tails of $A^{i}(E)$ in the unoccupied region of the
valence band (see \ref{Ghost}). For clarity, the PE spectra were evaluated
for the CFS mode, in which $k_{\bot }^{0}$ is constant (deviations from the
EDC\ mode are insignificant, because normally the final-state $\frac{%
\partial E^{f}}{\partial k_{\bot }}$ is large enough to ensure small
variations of $k_{\bot }^{0}$ through the peak profile).

The simulation was performed for four $\delta k_{\bot }$ values:\ zero for
the ideal $k_{\perp }$-resolution, $0.2$%
\mbox{$\vert$}%
$\Gamma X$%
\mbox{$\vert$}
as a representative for the BS-regime in a real PE experiment, $0.6$%
\mbox{$\vert$}%
$\Gamma X$%
\mbox{$\vert$}
for degraded $k_{\perp }$-resolution, and $2$%
\mbox{$\vert$}%
$\Gamma X$%
\mbox{$\vert$}
for the 1DOS-regime. The results are shown in Fig.3 as series of the PE
spectra with $k_{\bot }^{0}$ scanning along $\Gamma X$ (upper panels, each
series in different intensity scale) and positions of the spectral peaks
mapped on top of the true valence band (lower panels).

In the ideal $k_{\perp }$-resolution limit the PE peak reflects the
initial-state $A^{i}\left( E\right) $ and ideally follows the true $%
E^{i}(k_{\bot }^{0})$. Interestingly, upon passing $k_{\bot }^{F}$ there
appears an extremely small peak just below $E_{F}$; it replicates the
low-energy tail of the spectral function $A^{i}(E)$ whose maximum is already
above $E_{F}$ (see \ref{Ghost}).

In the BS-regime ($\delta k_{\bot }=0.2$%
\mbox{$\vert$}%
$\Gamma X$%
\mbox{$\vert$}
in our simulation) the $k_{\perp }$-resolution remains sufficient for the PE
peaks to closely follow $E^{i}(k_{\bot })$. This regime can be used for
accurate 3D band mapping. Two peculiarities should be noted here: 1) Despite 
$A^{i}(E)$ becomes the $\delta $-function upon approaching $E_{F}$, the PE\
peak does not sharpen up to singularity; 2) When $k_{\bot }^{0}$ passes $%
k_{\bot }^{F}$ and enters into the unoccupied region, the peaks do not
disappear but become 'ghost' peaks which are highly asymmetric and have
their maximum stationary in energy just below $E_{F}$. Both peculiarities
are appearances of the $k_{\perp }$-broadening and will be analyzed in \ref%
{Ghost} and \ref{Lineshapes}.

As the $k_{\perp }$-resolution degrades ($\delta k_{\bot }=0.6$%
\mbox{$\vert$}%
$\Gamma X$%
\mbox{$\vert$}%
), the concomitant averaging of $E^{i}(k_{\bot })$ causes asymmetry of the
peaks and strong {\em intrinsic shifts} of their maxima from the true $%
E^{i}(k_{\bot })$. Although the PE peaks deceptively keep dispersion on $%
k_{\bot }^{0}$, their use for 3D band mapping will return a distorted
valence band dispersion. Note appearance of an additional shoulder-like
structure, which originates from the large 1DOS in the bottom of the band
and becomes notable as the main peak shifts upwards.

In the 1DOS-regime ($\delta k_{\bot }=2$%
\mbox{$\vert$}%
$\Gamma X$%
\mbox{$\vert$}%
) the $k_{\perp }$-resolution is completely lost. The maxima of the PE\
peaks become stationary in energy reflecting the maximum of the valence band
1DOS (and a tiny replica of the $A^{i}(E)$ singularity at $E_{F}$, omitted
from the plot). The $k_{\bot }^{0}$ changes show up only in the intensity
modulation. Note however that the PE\ peaks are notably shifted from the
true 1DOS maximum.

The regimes of the PE experiment in general follow the well-known
''universal curve'' $\lambda (E)$ \cite{Kevan92,Hufner95,Feibelman74} which
gives the $k_{\bot }$-resolution as $\delta k_{\bot }=\lambda ^{-1}$. The
BS-regime holds while $\lambda $ remains large, which takes place at low $%
E^{f}$ below or not far above a plasmon excitation energy of 20-30 eV (see
examples in \ref{Non-linearity}). Upon further increase of energy $\lambda $
reaches its minimum around 50-100 eV, and the PE experiment enters into the
1DOS-regime, although in practice the $k_{\bot }$-dispersion signatures are
never completely suppressed. The BS-regime recovers in the high-energy
region above $\sim 300$ eV where $\lambda $ rises again (despite at these
energies the unoccupied bands become immense in number, only one of them
remains effective in the PE final state in the sense of effective coupling
to the outgoing plane wave \cite{CFC-Vg}).

The high-energy region is characterized by low PE intensity resulting from
small valence band crossection and from the Debye-Waller factor, in which
the increase of $h\nu $ reduces the PE intensity on equal footing with
temperature. However, the advent of new high-brilliance synchrotron
radiation sources and multidetection PE analysers has allowed for PE\
experiments in this region with reasonable count rates and resolution in $E$
and ${\bf k}_{\parallel }$ \cite{Saitoh00,Matsushita02}. ${\bf k}$-resolved
experiments were reported, for example, in \cite{Hoffmann02} where the $sp$%
-band of Al was found to be mirrored in the $h\nu $-dispersion of the PE
signal up to 750 eV. Interestingly, this study has found an increase of the
surface to bulk signal intensity ratio with $h\nu $, opposite to the
tendency expected from the ''universal curve'', and attributed this
surprising phenomenon to intrinsic phonon excitation effects involved in the
PE\ process. These results call for further studies on the role of the
phonon excitation in high-energy PE spectroscopy.

\section{Mechanisms limiting the intrinsic accuracy}

Degradation of the $k_{\bot }$-resolution, as we observed above, results in
intrinsic shifts of PE peaks from the true valence band $k_{\bot }$%
-dispersions, limiting the {\em intrinsic accuracy} of 3D band mapping.
These shifts can originate through various mechanisms, which will now be
analyzed.

\subsection{Non-linearity mechanism}

\label{Non-linearity}Degradation of the $k_{\bot }$-resolution can crawl
into the position of the PE peaks, firstly, by non-linearity of the valence
band $k_{\bot }$-dispersion. As evident from Fig.2, in this case the
integral number of the $dk_{\bot }$-intervals within the $k_{\bot }$%
-broadening will be different for the initial states whose $E^{i}$ is above
the direct-transition energy $E^{i}(k_{\bot }^{0})$ and for the states whose 
$E^{i}$ is below $E^{i}(k_{\bot }^{0})$, being larger where the 1DOS is
larger. Correspondingly, the PE peak becomes asymmetric. Its maximum
somewhat shifts towards larger integral number of $dk_{\bot }$, deviating
from the true $E^{i}(k_{\bot }^{0})$. In the band interior the PE peaks
shift towards the band edges, where the 1DOS is larger. Near the band edges
the peaks, on the opposite, shift towards the band interior (so-called
in-band shifting \cite{TMDCs,CFS}) because beyond the band edge there are no
states and the 1DOS is zero. We will refer to this mechanism to limit the
intrinsic accuracy as the {\em non-linearity} mechanism. It was behind the
intrinsic shifts observed in the PE simulation in Fig.3, where the PE peaks
clearly followed the characteristic pattern of shifting towards larger 1DOS.
Note that the intrinsic shifts due to the non-linearity mechanism depend,
apart from $\delta k_{\bot }$ and non-linearity of $E^{i}(k_{\bot })$, on
the initial-state $\delta E$.

The non-linearity mechanism can be illustrated by experimental data for VSe$%
_{2}$ (for a detailed discussion see \cite{TMDCs}). Information about the
final-state dispersions and lifetimes, required for determination of $\delta
k_{\bot }$, was achieved in this case by VLEED (see \ref{Optim}). The
experimental energy dependence of $V_{i}$ is shown in Fig. 4 ({\it a}). It
is characterized by a sharp increases of $V_{i}$ at the plasmon excitation
energy $\hbar \omega _{p}$. The experimental $E^{f}(k_{\bot })$ is shown in
Fig. 4 ({\it b}). Note that these final band can hardly be described by
free-electron-like dispersions: there are {\em two} bands, and each can be
free-electron fitted only locally with the inner potential strongly
depending on $E^{f}$ and ${\bf k}_{\parallel }$ (such a complicated
structure of the final states is in fact typical of many materials with
large unit cell, containing more than one atom, due to backfolding of many
free-electron bands into the first BZ and their mutual hybridization). An
estimate of $\delta k_{\bot }$ was obtained, neglecting some increase in the
final-state band gaps, as $2V_{i}\left( \frac{\partial E^{f}}{\partial
k_{\bot }}\right) ^{-1}$ (see \ref{Optim}). The obtained $\delta k_{\bot }$
values are shown in Fig. 4 ({\it b}) superimposed on the final bands.
Following the increase of $V_{i}$, the $k_{\bot }$-resolution degrades above 
$\hbar \omega _{p}$. The results of PE band mapping, the binding energy of
the PE peaks as a function of $k_{\bot }^{0}$, are shown in Fig. 4 ({\it c})
with distinction for the final states below and above $\hbar \omega _{p}$.
Near the band edges of the 3D bands the experimental points above $\hbar
\omega _{p}$ are systematically shifted towards the band interior compared
to those below $\hbar \omega _{p}$. It should be stressed that here we have
an explicit example how {\em the same} initial states yield {\em different}
energies of the PE peaks depending on the $k_{\bot }$-resolution. The
observed shifts demonstrate the intrinsic effect of in-band shifting,
occurring mainly by the non-linearity mechanism. Interestingly, increase of $%
E^{f}$ towards 50 eV results in further increase of $\delta k_{\bot }$ and
suppression of any dispersion of the PE peaks \cite{Starnberg94}.

Our analysis demonstrates that for VSe$_{2}$ the energy region of accurate
3D band mapping is exhausted by low $E^{f}$ below or not far above $\hbar
\omega _{p}$. Such a situation is in fact fairly general, because a
pronounced plasmon threshold is typical of many materials with some
exceptions such as Cu and Ni. The intrinsic accuracy effects similar to VSe$%
_{2}$ were also observed for TiS$_{2}$ \cite{TMDCs}, WSe$_{2}$ \cite%
{Finteis97}, Ge \cite{Grobman75}, etc.

In the final-state band gaps the Bloch waves experience additional damping
due to elastic scattering from the crystal potential \cite{CFS,Courths89}.
Although this is normally much weaker than the damping due to $V_{i}$, as in
the above example, for materials with exceptionally large final-state band
gaps the additional damping can cause notable degradation of the $k_{\bot }$%
-resolution. Graphite is a typical example of this (for a detailed
discussion see \cite{Graphi,GraphiPE}). Its final bands were simulated using
the empirical pseudopotential method based on the VLEED data for their
energies in the $\Gamma $-point (the simulation is therefore less accurate
in the BZ interior) and $V_{i}$. The resulting $E^{f}(k_{\bot })$ is shown
in Fig.5 ({\it a}) and ({\it b}) as a function of $Imk_{\bot }^{f}$ and $%
Rek_{\bot }^{f}$, respectively, using the double-BZ representation to
reflect the dipole selection rules in graphite. The final states are
characterized by a strong increase of $\delta k_{\bot }=2Imk_{\bot }^{f}$ in
the two band gaps in the $\Gamma $-point. Based on these data, we simulated
the PE response of the valence $\pi $-band, having the 3D character, using
the same formalism (\ref{IntF}). The results are shown in Fig.5 ({\it b}).
When the final-state $k_{\bot }^{0}$ moves towards the $\Gamma $-point and
enters the band gaps, progressive degradation of the $k_{\bot }$-resolution
results, by the non-linearity mechanism, in increasing of the in-band
shifts. In the lower band gap, which has the maximal width, the increase of $%
\delta k_{\bot }$ is particularly strong; the concomitant in-band shifting
is so dramatic that it overcomes the trend dictated by the band dispersions
themselves and {\em reverses} the dispersion of the PE peaks compared to the
true $\pi $-band. Such an unusual reversed dispersion has indeed been found
in the PE experiment \cite{GraphiPE}.

Typically, intrinsic shifts near the band extrema due to the non-linearity
mechanism are of the order of a few tenths of eV. The experiment often
suggests however larger shifts \cite{TMDCs,Starnberg94,Finteis97}. As
degradation of the $k_{\bot }$-resolution goes along with reduction of the
escape length $\lambda $, this reveals certain contribution of surface
effects such as suppression of the 1DOS in the band extrema (see \ref{Other}%
). It should be noted that the condition (\ref{BSRegime}) identifying the
BS-regime is particularly restrictive for layered materials owing to their
small $k_{\bot }^{BZ}$ ( see \cite{Kluwer} for a specialized survey).

\subsection{'Ghost' peaks at $E_{F}$}

\label{Ghost}An interesting phenomenon occurs when a 3D band crosses the
Fermi surface. This situation is described by the PE simulation in Fig.3:
When the direct-transition $k_{\bot }^{0}$ moves outside the Fermi surface
and starts sampling the unoccupied part of the valence band, the PE peaks,
mysteriously, do not disappear. Instead, they become asymmetric and
stationary in energy just below the Fermi level $E_{F}$, and only gradually
reduce in amplitude. Experimentally, such a phenomenon has been observed,
for example, in a famous PE study on Na, where the $sp$-band crosses $E_{F}$ %
\cite{Jensen85,Shung86}. We will refer to this phenomenon as {\em 'ghost'
peaks}. It has in fact two contributions due to different mechanisms.

The first contribution is due to a general mechanism taking place for 3D as
well as 2D bands. It comes from the {\em unoccupied} states next to $E_{F}$:
The spectral function $A^{i}(E)$, centered at the band energy above $E_{F}$,
protrudes its low-energy tail, however small the remaining amplitude, below $%
E_{F}$ and gives therefore some PE intensity (such an effect for 2D bands is
discussed, for example, in \cite{Claessen92,Claessen96}). Such a
contribution can be isolated in our PE simulation by restricting the
integration (\ref{IntF}) to $k_{\bot }>k_{\bot }^{F}$. It survives even in
the limit of the ideal $k_{\perp }$-resolution, see Fig.3. In any case, the
'ghost' peak intensity due this mechanism is very small.

The second contribution, much predominating for the 3D bands, comes from the
occupied states: As illustrated in Fig.6, even if $k_{\bot }^{0}$ moves
outside the Fermi surface, the $dk_{\bot }$-states at the initial-state $%
E^{i}$ below $E_{F}$ give certain 'ghost' intensity by virtue of being
accessible via the $k_{\perp }$-broadening. Upon moving $k_{\bot }^{0}$
further from $k_{\bot }^{F}$ the $A^{f}\left( k_{\bot }\right) $ profile
shifts, reducing its amplitude at the initial band and, concomitanly, the
'ghost' peak amplitude. Note that this mechanism gives a significant 'ghost'
peak intensity even if the initial-state $E$-broadening is vanishing which
takes place upon approaching $E_{F}$.

The 'ghost' peak phenomenon complicates identification of the $E_{F}$
crossings in mapping of the Fermi surface formed by 3D bands. A well-known
example is NbSe$_{2}$, whose 4{\it p}$_{z}$ band forms a 3D pocket of the
Fermi surface with extension in $k_{\bot }$ small compared to the $k_{\bot }$%
-broadening; direct mapping of this pocket is precluded by non-dispersive
'ghost' intensity \cite{NbSe2,Rossnagel01}. Even if the mapping is performed
as a function of ${\bf k}_{\parallel }$, the 'ghost' peaks stay at $E_{F}$
through the entire interval of ${\bf k}_{\parallel }$ where at least some
part of the 3D band remains occupied. It should be noted that often the PE
data, especially for quasi-2D materials, is interpreted neglecting the 3D
effects; in this case such 'ghost' peaks in the ${\bf k}_{\parallel }$%
-dispersion are misinterpreted as a sign of unusual self-energy effects. A
way to circumvent the 'ghost' peak problem in the Fermi surface mapping was
suggested in \cite{CFS}: one varies ${\bf k}_{\parallel }$, simultaneously
changing $E^{f}$ to remain in the $E^{i}\left( k_{\bot }\right) $ minimum.

\subsection{Linewidths near $E_{F}$}

\label{Lineshapes}Similar in origin to the 'ghost' peaks is an interesting
PE linewidth behavior upon approaching $E_{F}$: As seen in our simulations
in Fig.3 for the BS-regime, the PE peak from the 3D band, mysteriously,
retains a finite linewidth up to $E_{F}$ without sharpening up as might be
expected from the behavior of the initial-state $A^{i}(E)$, which becomes
singular at $E_{F}$ because of the vanishing $E$-broadening. The origin of
such a counter-intuitive linewidth behavior is again due to the 3D character
of the band and the $k_{\perp }$-broadening \cite{Smith93,Smith92}: As one
can conjecture from Fig.6, even if the $E$-broadening is zero, the initial
states around the direct-transition energy $E^{i}(k_{\bot }^{0})$ are
accessible via the $k_{\perp }$-broadening and contribute to the PE peak
width. In the ideal $k_{\perp }$-resolution limit, as demonstrates our
simulation in Fig.3, the PE peak well becomes singular towards $E_{F}$.

An experimental example of such an effect of the $k_{\perp }$-broadening on
the linewidths can be seen, for example, in the PE peak from the $sp$-band
of Cu whose linewidth increases upon approaching $E_{F}$ \cite{Berge02}. In %
\cite{Smith93,Smith92} it was shown that the same effect can explain an
unusual linewidth energy dependence near $E_{F}$ observed in Bi$_{2}$Sr$_{2}$%
CaCu$_{2}$O$_{8}$, which otherwise would indicate deviations from the Fermi
liquid picture.

Significant PE intensity next to $E_{F}$, seen in all regimes of the
simulation in Fig.3, also originates from existence of a 3D band crossing $%
E_{F}$ and the $k_{\bot }$-broadening. Such a peculiarity is typical of the
PE experiments on 3D materials such as Cu.

\subsection{Other mechanisms}

\label{Other}Intrinsic shifts in the 3D bands can also appear due to a {\em %
matrix element} mechanism: If $M_{fi}$ as a function of energy and $k_{\perp
}$ undergoes sharp variations within the $k_{\bot }$-broadening, a
contribution of the states above and below $E^{i}(k_{\bot }^{0})$ is
different (see Fig.2). The peak becomes asymmetric and it maximum
experiences an intrinsic shift from the true band energy. Such a mechanism
has been identified, for example, in TiS$_{2}$ \cite{TMDCs}. Similar effects
due to $T^{f}$ should be less pronounced, because the energy variations of $%
T^{f}$ are normally smooth and become sharper only at extremely low $E^{f}$,
as can be seen in VLEED data (see, e.g., \cite{TMDCs,Graphi,Krasovskii02})

The above mechanisms were all based on the bulk bandstructure picture, in
which the bandstructure near the surface is considered identical to the bulk
one. This implies that the crystal potential, and thus the wavefunctions,
ideally repeat the bulk ones up to the surface. In fact, within a few atomic
layers of the PE\ escape depth they can experience notable modification.
Such {\em surface effects} also give rise to intrinsic shifts. For example,
the 1DOS singularities in the band extrema, characteristic of the bulk $E(%
{\bf k})$, are smeared in the local DOS\ near the surface (see, e.g., \cite%
{Pehlke87,Fang97,Courths01}), which results in an increase of the in-band
shifting. The surface effects are harder to control compared to the $k_{\bot
}$-resolution effects. Qualitatively, their contribution reduces with
increase of $\lambda $ simultaneously with reduction of the intrinsic shifts
of the bulk origin due to the concomitant improvement of the $k_{\bot }$%
-resolution.

{\em Surface photoelectric effect}, generated by abrupt change in the
dielectric response at the surface barrier and described by the ${\bf %
\bigtriangledown }\cdot {\bf A}$ part of $M_{fi}$, can influence the PE
signal from the 3D bands through interference of the surface and bulk PE
components (see, e.g., \cite{Levinson79,Hansen97,Michalke00}). Normally this
effect only causes some asymmetry of the PE peaks. However, near the surface
and bulk plasmon excitation energies, especially for free-electron metals
such as Na, the lineshape changes can be drastic and return large intrinsic
shifts \cite{Shung86,Claesson99}. In \cite{Michalke00} it was demonstrated
that the surface and bulk components can be experimentally separated by
changing the light incidence angle.

\subsection{Intrinsic accuracy within the one-step PE theory}

\label{OneStep}The intrinsic effects are naturally embodied in the one-step
PE theory (see, e.g., \cite{Feibelman74,Pendry76}) which finds the PE
intensity as a matrix element $\left| \left\langle \Phi ^{f}\left| \widehat{%
{\bf A}}\cdot \widehat{{\bf p}}\right| \Phi ^{i}\right\rangle \right| $
between the final- and initial-state wavefunctions $\Phi ^{f}$ and $\Phi
^{i} $. Interference between different Bloch wave constituents in the final
and initial states, matrix element effects and surface effects, all ignored
in the simplified formalism (\ref{IntF}) used above, are naturally included
here. However, a heavy computational machinery often obscures the physical
mechanisms.

An example of the intrinsic accuracy analysis using the one-step PE theory
can be found in \cite{Pehlke89} in application to TiSe$_{2}$. Similarly to
the above results for VSe$_{2}$, for the final states above $h\omega _{p}$
the $k_{\bot }$-resolution was found to degrade, causing strong broadening
and intrinsic shifts of the PE peaks. Surface effects due to smearing of the
bulk 1DOS singularities have also been identified. Another example is a
calculation on Na \cite{Shung86}, which described the 'ghost' peaks at $%
E_{F} $ and in-band shifting in the bottom of the valence band.

\section{Optimization of the PE experiment using VLEED}

\label{Optim}3D band mapping remains relevant if the intrinsic shifts are
small. Of the factors affecting these shifts, the non-linearity of the $%
k_{\bot }$-dispersion, initial-state $E$-broadening and, partly, variations
of $M_{fi}$ are all due to inherent properties of the valence bands under
study and remain beyond control by the experimentator. The $k_{\perp }$%
-resolution, on the contrary, is an exclusively final-state property, and by
tuning $E^{f}$ can taken to values which ensure negligible intrinsic shifts.
This is achieved in the BS-regime $\delta k_{\bot }\ll k_{\bot }^{BZ}$ (see %
\ref{Regimes}). The exact energy ranges of this regime are however much
material dependent. Knowledge of the $k_{\bot }$-resolution energy
dependence is therefore required.

VLEED, covering the energy range below $\sim $40 eV, has recently been
established as the experimental method giving the most direct access to the
PE final states with resolution in the 3D wavevector (see \cite%
{Exi,TMDCs,CFS} and references therein). In the VLEED spectra of elastic
electron reflection, the energies of the spectral structures give the
critical points in the final bands. $E^{f}(k_{\bot })$, obtained from this
experimental data, can further be used to control $k_{\perp }$ in the PE
experiment. Broadening of the VLEED structures, moreover, gives an estimate
of the corresponding $V_{i}$ values \cite{TMDCs,Krasovskii02,Barrett03}. By
performing calculations of the complex bandstructure with the experimental $%
V_{i}$ \cite{Exi,CFS,Krasovskii02}, one obtains the $k_{\bot }$-resolution
as $\delta k_{\bot }=2Imk_{\bot }^{f}$ of the generated Bloch waves (and
simultaneously the energy dependence of $\lambda $, as it was recently
demonstrated for graphite in comparison with determination of $\lambda $ by
the conventional overlayer method \cite{Barrett03}). As such calculations
can however be time-consuming, $\delta k_{\bot }$ can be estimated in a
simplified manner directly from the $V_{i}$ values and the final-state group
velocity $\frac{\partial E^{f}}{\partial k_{\bot }}$ as \cite%
{Pendry69,Knapp79,Smith93}%
\begin{equation}
\delta k_{\bot }\sim 2V_{i}\left( \frac{\partial E^{f}}{\partial k_{\bot }}%
\right) ^{-1}
\end{equation}%
This estimate is accurate however only outside the final-state band gaps.
Nevertheless, increase of$\ \delta k_{\bot }$ associated with additional
damping in the gaps (see Fig.5) is normally less significant, except for a
few materials with exceptionally wide band gaps such as graphite.

Use of the VLEED predictions on the $k_{\bot }$-resolution allows the
experimentator to {\em optimize} the PE experiment on the intrinsic accuracy
by selecting the $E^{f}$ intervals which ensure the BS-regime (see the
examples in \ref{Non-linearity}).

As a concluding remark, it should be noted that one can significantly
improve accuracy of 3D band mapping by computational modelling of the
intrinsic accuracy effects, as described above, and correcting the
experimental data correspondingly. This is often critical for proper
evaluation of the true excited-state self-energy effects, for example in
graphite \cite{GraphiPE} and Na \cite{Shung86}.

\section{Conclusion}

Fundamental principles of PE have been analysed, focussing on emission from
3D\ states in the valence band. Damping of the final state due to inelastic
and elastic scattering in the crystal results in equivalent broadening in $%
k_{\bot }$, which acts as the intrinsic $k_{\bot }$-resolution of the PE
experiment. The PE peak is then formed as a matrix-element weighted average
of the quasiparticle $k_{\bot }$-dispersion in valence band, which can be
shifted from the dispersion itself. Such shifts intrinsically limit the
accuracy of 3D band mapping, which remains accurate only in the
bandstructure regime, characterized by the intrinsic $k_{\bot }$-broadening
small compared to the BZ extension in $k_{\bot }$. Upon degradation of the $%
k_{\bot }$-resolution the PE experiment yields increasing intrinsic shifts
and, finally, enters into the 1DOS regime. The intrinsic $k_{\bot }$%
-resolution can be controlled using VLEED, which gives the final-state
dispersions and lifetimes. The mechanisms forming the intrinsic shifts have
been exposed, including non-linearity of the $k_{\bot }$-dispersion,
variations of the matrix element and surface effects. Appearance of the
intrinsic shifts and transformation of the regimes of the PE experiment has
been analysed using a physically transparent picture of interplay of the
final- and initial-state spectral functions, and illustrated by PE
simulations and experimental examples. The $k_{\perp }$-broadening gives
rise to other surprising effects typical of 3D valence states such as
'ghost' peaks whose direct-transition $k_{\perp }$ falls outside the Fermi
surface, and finite width of the PE peaks at the Fermi level.

\bigskip

I thank R. Claessen, P.O. Nilsson and R. Feder for promoting discussions.
Financial support by the Deutsche Forschungsgemeinschaft (CL124/5-1) is
gratefully acknowledged.

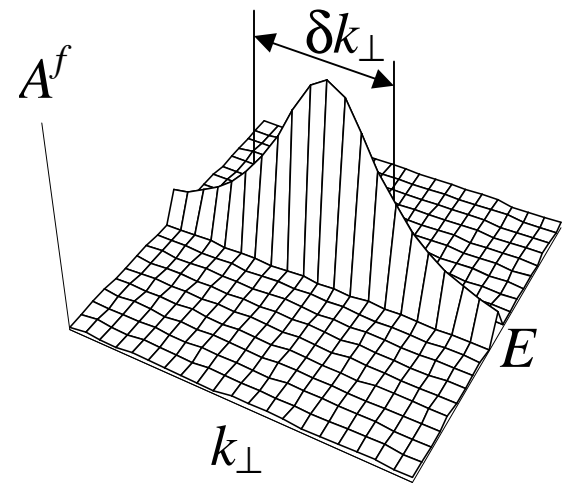
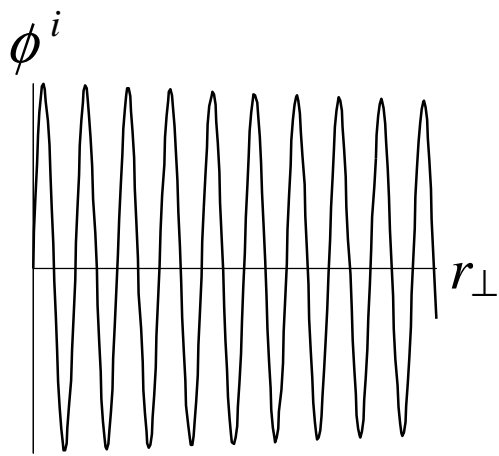
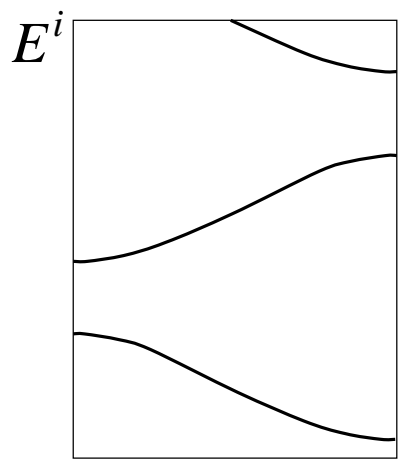
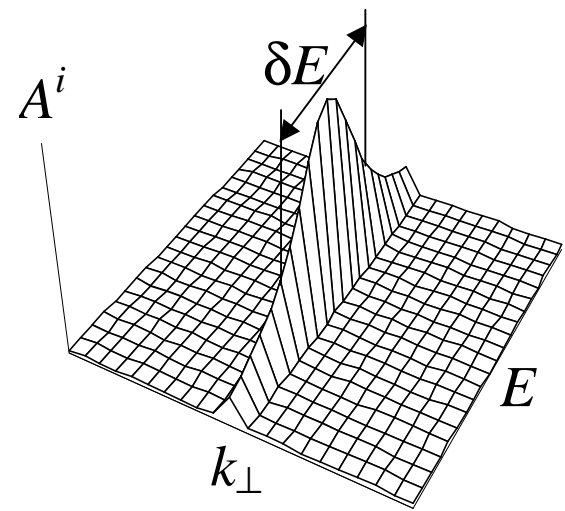
\begin{figure}[tbp]
\caption{Characteristic behavior of wavefunctions, $k_{\bot }$-dispersions
and spectral functions of the final and initial states of PE, shown,
respectively, on the upper and lower panels. The final states are
characterized by fixed $E$ and broadening in $k_{\bot }$, while the initial
state, complementarily, by fixed $k_{\bot }$ and broadening in $E$.}
\end{figure}

\begin{figure}[tbp]
\caption{Development of the PE peak. Its broadening results from the
final-state $k_{\bot }$-broadening, defining the intrinsic $k_{\bot }$%
-resolution $\protect\delta k_{\bot }$, combined with the initial-state $E$%
-broadening $\protect\delta E$. The maximum of the peak shows an intrinsic
shift towards the larger average 1DOS from the position $E^{i}(k_{\bot
}^{0}) $ dictated by the direct transition at $k_{\bot }^{0}$.}
\end{figure}

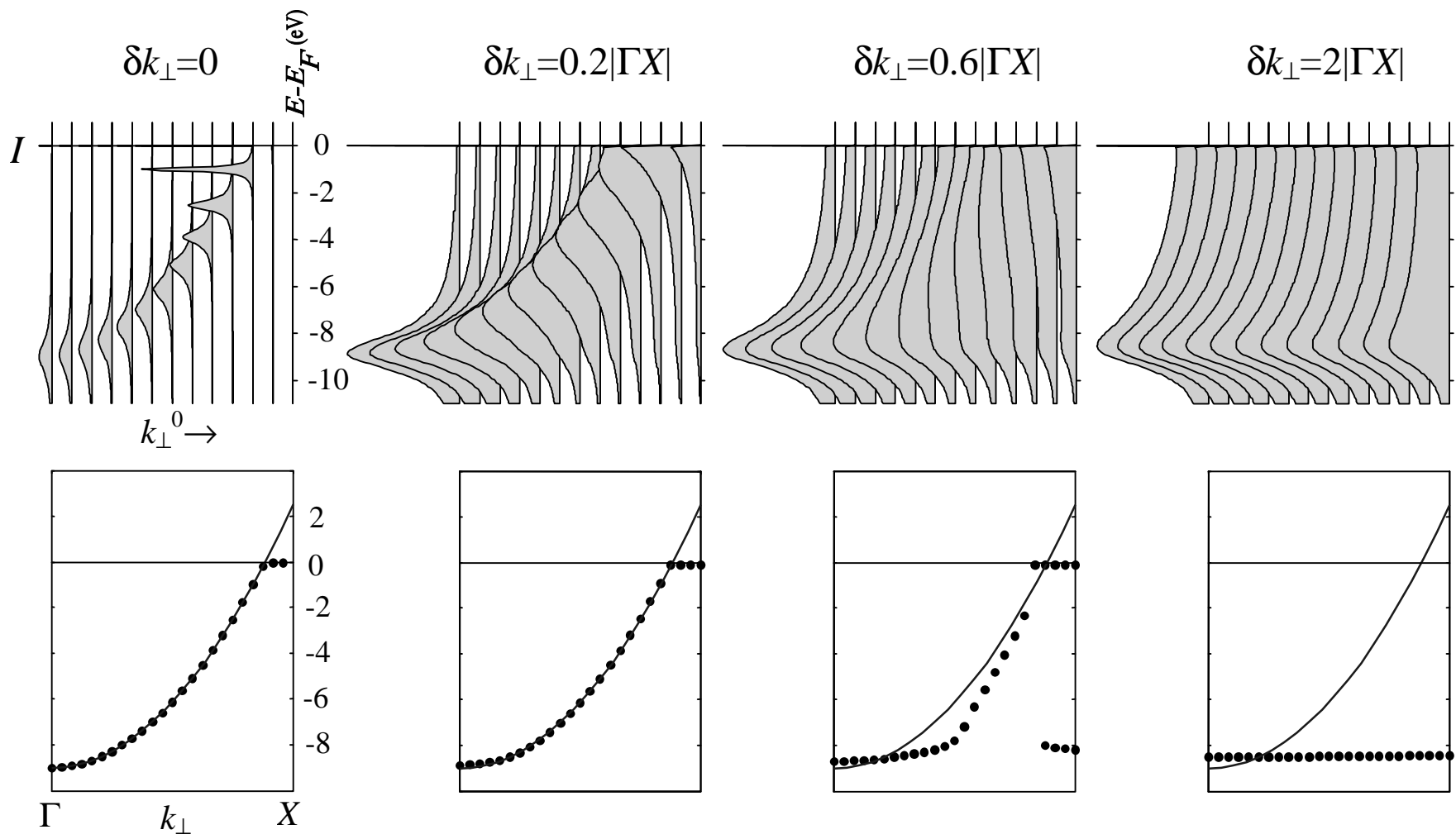
\begin{figure}[tbp]
\caption{Simulated PE\ experiments on a model 3D band, resembling the $sp$%
-band of Cu, performed with different strength of damping in the final
state. The corresponding $\protect\delta k_{\bot }$ values of the intrinsic
the $k_{\perp }$-resolution (indicated on {\it top}) are characteristic of
the BS-regime ({\it left}), degraded $k_{\perp }$-resolution ({\it center}),
and 1DOS-regime ({\it right}). The results are shown as series of a few PE
spectra ({\it upper panels}) whose offsets reflect scanning of the
direct-transition $k_{\bot }^{0}$ along $\Gamma X$, and the position of the
spectral peak maxima on top of the true valence band $E^{i}(k_{\bot })$ (%
{\it lower panels}). Finite width of the PE peak upon approaching $E_{F}$
and 'ghost' peaks remaining after the $E_{F}$ crossing as both effects of
the $k_{\bot }$-broadening. Degradation of the $k_{\perp }$-resolution
results in intrinsic shifts of the PE peaks, in this case is due to the
non-linearity mechanism, from the true valence band.}
\end{figure}

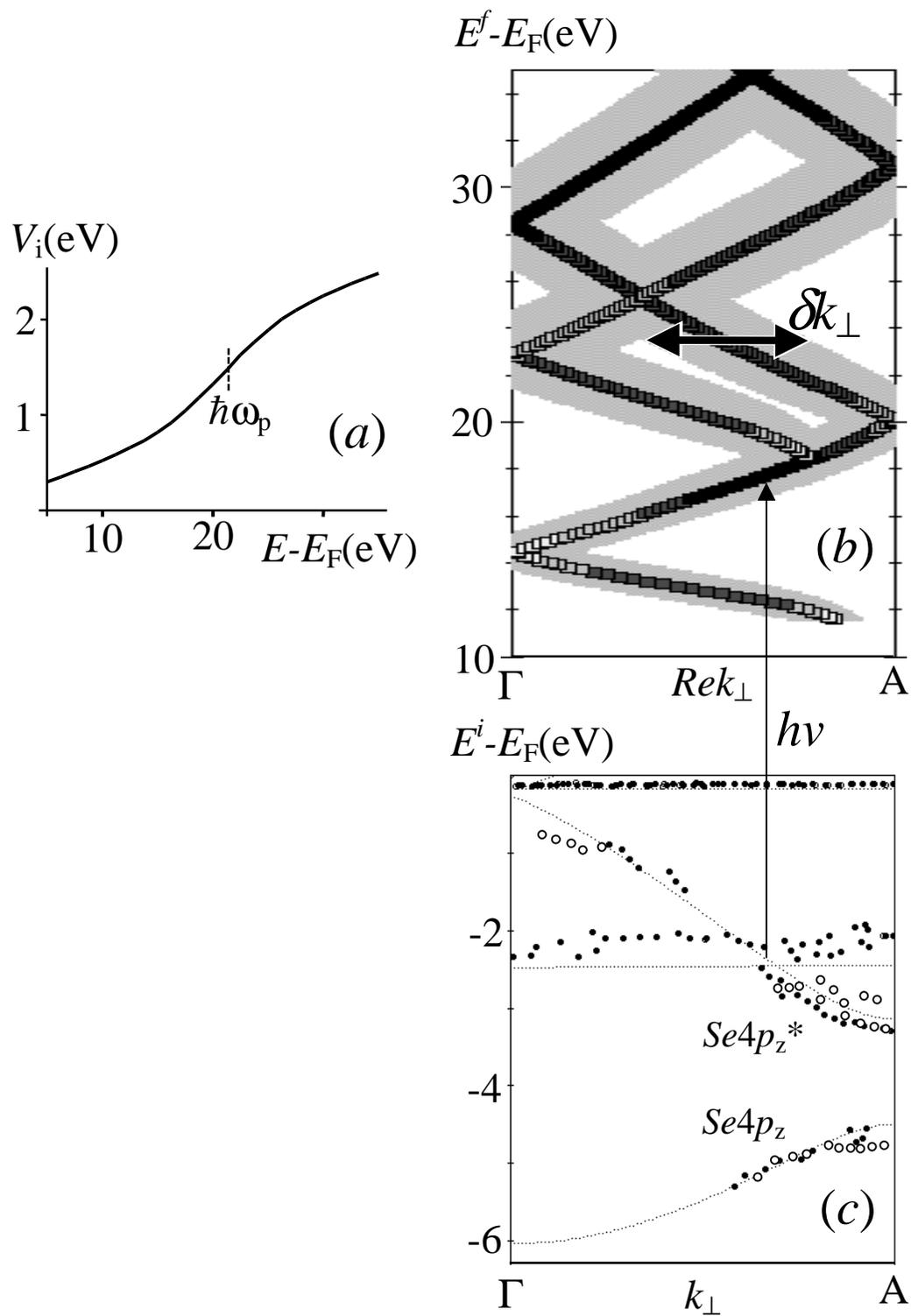
\begin{figure}[tbp]
\caption{Intrinsic accuracy for VSe$_{2}$: ({\it a}) Energy dependence of
the final-state $V_{i}$ with a characteristic increase due to the plasmon
excitation at $\hbar \protect\omega _{p}$; ({\it b}) Final-state bands
(including, of all unoccupied bands, only those whose Bloch wave effectively
couples to the outgoing photoelectron plane wave). The $k_{\bot }$%
-resolution is shown by shading; ({\it c}) Position of the PE peaks
superimposed on DFT-LAPW calculations. The points obtained with $E^{f}$
below $\hbar \protect\omega _{p}$ are shown as solid dots, and above as open
dots. A systematic in-band shifting in the 3D bands, following degradation
of the $k_{\bot }$-resolution above $\hbar \protect\omega _{p}$, is mainly
due to the non-linearity mechanism.}
\end{figure}

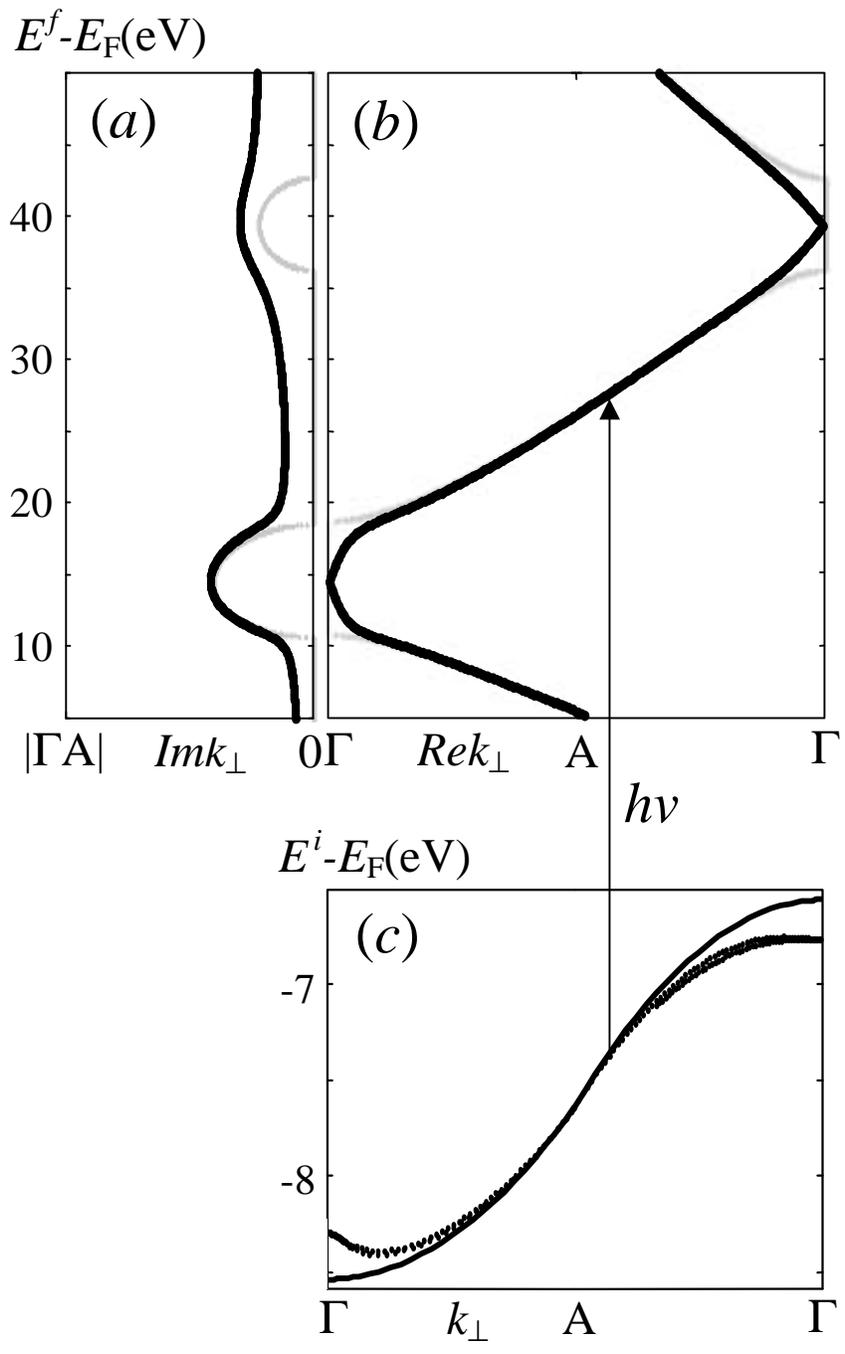
\begin{figure}[tbp]
\caption{Intrinsic accuracy for graphite: ({\it a,b}) Empirical
pseudopotential simulation of the final bands based on VLEED data (black
lines); the band gaps are emphasized by setting $V_{i}$ to zero (gray); (%
{\it c}) Simulated position of the PE\ peaks (dense dots) from the valence $%
\protect\pi $-band on top of its true dispersion. Severe degradation of the $%
k_{\bot }$-resolution $\protect\delta k_{\bot }=Imk_{\bot }^{f}$ in the
final-state band gaps causes large in-band shifts due to the non-linearity
mechanism, and even reverses the dispersion of the PE peaks in the lower
gap. }
\end{figure}

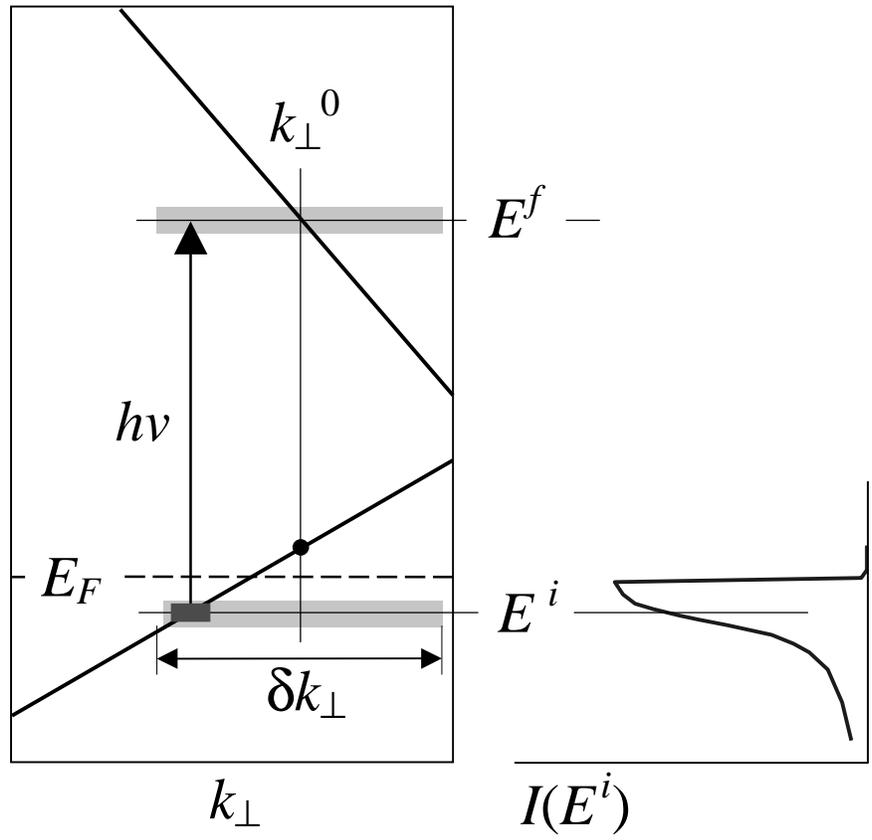
\begin{figure}[tbp]
\caption{Mechanism of the 'ghost' peaks from 3D bands crossing the Fermi
surface. Even if the direct-transition $k_{\bot }^{0}$ is in the unoccupied
region, the PE signal comes from the occupied part of the valence band
accessible via the final-state $k_{\perp }$-broadening.}
\end{figure}

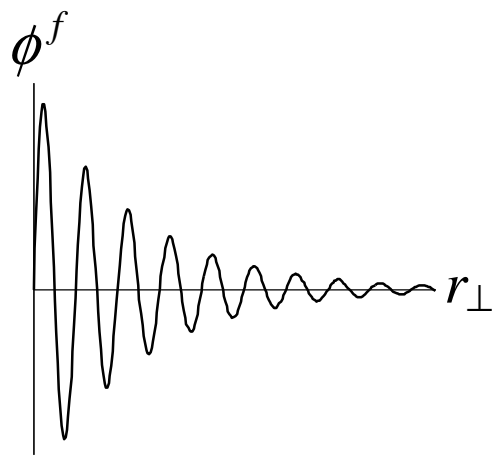
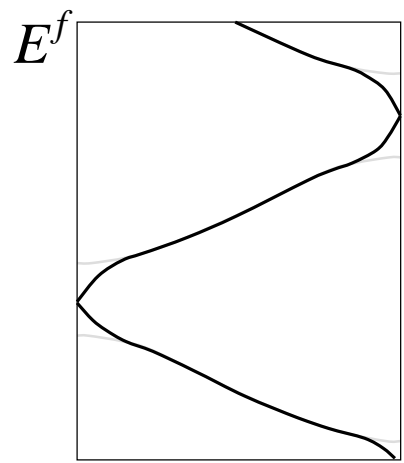
\end{document}